\documentclass[twocolumn,showpacs,preprintnumbers,amsmath,amssymb]{revtex4}

\usepackage{graphicx}
\usepackage{dcolumn}
\usepackage{bm}

\begin{document}

\title{Layered Structures Favor Superconductivity in Compressed Solid SiH$_{4}$}

\author{X. J. Chen$^{1,2,3}$, J. L. Wang$^{1}$, V. V. Struzhkin$^{2}$, H. K. Mao$^{2}$,
R. J. Hemley,$^{2}$ and H. Q. Lin$^{1}$ }

\affiliation{$^{1}$Department of Physics and Institute of
Theoretical Physics,
Chinese University of Hong Kong, Hong Kong, China \\
$^{2}$Geophysical Laboratory, Carnegie Institution of Washington,
Washington, DC 20015, USA \\
$^{3}$School of Physics, South China University of Technology,
Guangzhou 510640, China}

\begin{abstract}
The electronic and lattice dynamical properties of compressed solid
SiH$_{4}$ have been calculated in the pressure range up to 300 GPa
with density functional theory. We find that structures having a
layered network with eight-fold SiH$_{8}$ coordination favor
metallization and superconductivity. SiH$_{4}$ in these layered
structures is predicted to have superconducting transition
temperatures ranging from 20 to 80 K, thus presenting new
possibilities for exploring high temperature superconductivity in
this hydrogen-rich system.
\end{abstract}
\date{\today}
\maketitle

SiH$_{4}$ offers a source for high purity silicon in epitaxial and
thin film deposition which is at the base of electronics and
microdevices. There is ongoing interest in this material as well due
to the suggestion of Ashcroft \cite{ashc} that SiH$_{4}$ would
eventually undergo a transition to metallic and then a
superconducting state at pressures considerably lower than may be
necessary for solid hydrogen. Exploring the possibility of metallic
hydrogen has long been a major driving force in high-pressure
condensed matter science and remains an important challenge in
modern physics and astrophysics. Recent experimental work on
SiH$_{4}$, using diamond-anvil cell techniques, has revealed an
enhanced reflectivity with increasing pressure \cite{sun,chen}. It
was found \cite{chen} that solid SiH$_{4}$ becomes opaque at 27-30
GPa and exhibits Drude-like behavior at around 60 GPa, signalling
the onset of pressure-induced metallization. Structural information
is the primary step toward understanding these observed electronic
properties.

SiH$_{4}$ has a rich phase diagram with at least seven known phases
\cite{chen,clus,wild}. Only one solid phase has been reported in the
pressure range between 10 and 25 GPa and at room temperature, with a
monoclinic structure (Space group $P2_{1}/c$) \cite{degt}. The very
low hydrogen scattering cross section of hydrogen-containing
materials in all diffraction methods makes structural determination
very difficult, specifically in determining the H positions.
Although the neutron diffraction is powerful in detecting the
H-bonding structures, the current accessible pressure range of this
technique is limited to 30 GPa \cite{ding}. Therefore it is still
impossible to use neutron diffraction to obtain the interesting
structural information of SiH$_{4}$ in the metallic state. The
challenge of experimentally or theoretically determining the
high-pressure structures of SiH$_{4}$ is still enormous.

The sequence of SiH$_{4}$ structures provides the basis for
understanding whether the material is a favorable candidate of a
high temperature superconductor. For a metallic $Pman$ SiH$_{4}$
phase, Feng {\it et al.} \cite{feng} obtained a superconducting
transition temperature $T_{c}$ of 166 K at 202 GPa by using the
electron-phonon coupling strength for lead under ambient pressure as
both materials have the same characteristic density of states per
volume at relevant pressures. Pickard and Needs \cite{pick} also
studied the structural properties of SiH$_{4}$ and mentioned the
possibility of superconductivity in a $C2/c$ phase. All previous
work \cite{degt,feng,pick} was done without including the
calculations of the phonon spectra and electron-phonon coupling
parameters. Later phonon calculations by Yao $et$ $al.$ \cite{yao}
showed that the $Pman$ structure is in reality not stable and that a
new $C2/c$ structure is dynamically stable from 65 to 150 GPa. This
$C2/c$ SiH$_{4}$ phase was predicted to exhibit superconductivity
close to 50 K at 125 GPa \cite{yao}. It is not known whether there
exists a common structural feature that favors superconductivity in
metallic SiH$_{4}$, and no study on superconductivity with other
stable structures has been attempted.

In this Letter we report a theoretical study of superconductivity in
compressed solid SiH$_{4}$ including structural, electronic, and
vibrational calculations. We find six energetically favorable
structures in which the $P\overline{1}$, $Cmca$, and $C2/c$
structures have layered networks with eight-fold SiH$_{8}$
coordination. This layered feature favors metallization and
superconductivity. The layered phases are predicted to have
$T_{c}$'s in the range of 20 and 80 K, suggesting that SiH$_{4}$ is
indeed a good candidate for high-temperature superconductivity.

To study the structural and electronic behavior of SiH$_{4}$ over a
wide range of pressure, we used the Perdew-Burke-Ernzerhof
generalized gradient approximation (GGA) density functional and
projector augmented wave method as implemented in the Vienna $ab$
$initio$ simulation package (VASP) \cite{vasp}. An energy cutoff of
450 eV is used for the plane wave basis sets, and
16$\times$16$\times$16 and 8$\times$8$\times$8 Monkhorst-Pack
$k$-point grids are used for Brillouin zone sampling of two SiH$_4$
molecular cells and four SiH$_4$ molecular cells, respectively. The
lattice dynamical and superconducting properties were calculated by
the Quantum-Espresso package \cite{pwscf} using Vanderbilt-type
ultrasoft potentials with a cut-off energy of 25 Ry and 200 Ry for
the wave functions and the charge density, respectively.
12$\times$12$\times$12 Monkhorst-Pack $k$-point grids with Gaussian
smearing of 0.02 Ry were used for the phonon calculations at
4$\times$4$\times$4 $q$-point mesh and double $k$-point grids were
used for calculation of the electron-phonon interaction matrix
element.

\begin{figure}[tbp]
\includegraphics[width=\columnwidth]{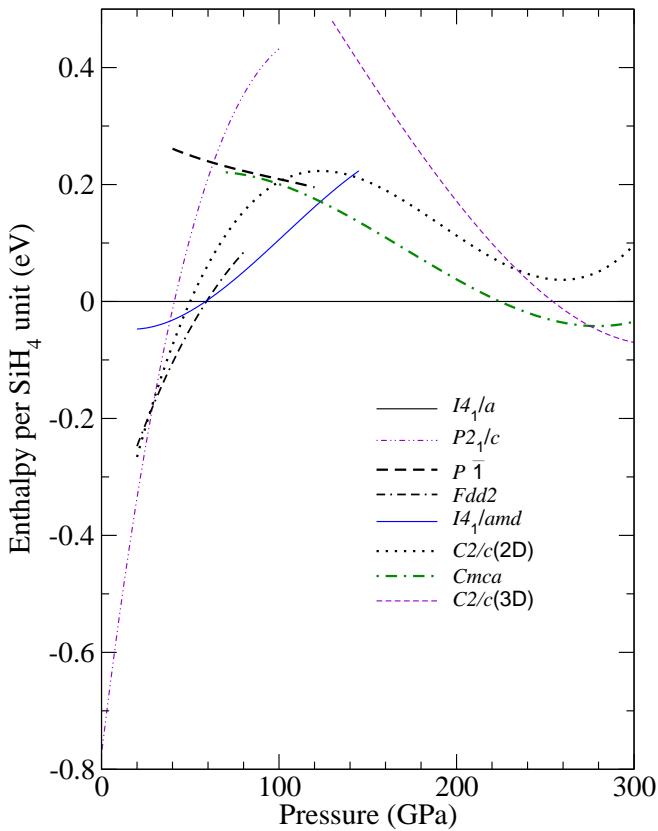}
\caption{(color online) The enthalpy versus pressure for competitive
structures of SiH$_{4}$. The enthalpy of the $I4_{1}/a$ phase is
taken as the reference point. }
\end{figure}

We performed a systematic study of the phase stability of SiH$_{4}$
based on {\it ab initio} first-principles calculations. Out of more
than one hundred structures we studied, six new polymorphs of
SiH$_{4}$ with low enthalpies are found in the pressure range from 0
to 300 GPa. In Fig. 1 we plot the pressure dependence of their
enthalpies along with the results for the $C2/c$ and $I4_1/a$
structure reported previously \cite{pick}. A monoclinic structure
with $P2_{1}/c$ symmetry has the lowest enthalpy below 27 GPa, in
good agreement with the recent experiments \cite{degt}. The $P2_1/c$
structure consists of four isolated covalently bonded SiH$_4$
tetrahedra with the H atom of one molecule pointing away from the H
atoms of a neighboring molecule. We find that a face-centered
orthorhombic structure with $Fdd2$ symmetry appears stable from 27
to 60 GPa. These two phases are insulating.

\begin{figure}[tbp]
\includegraphics[width=\columnwidth]{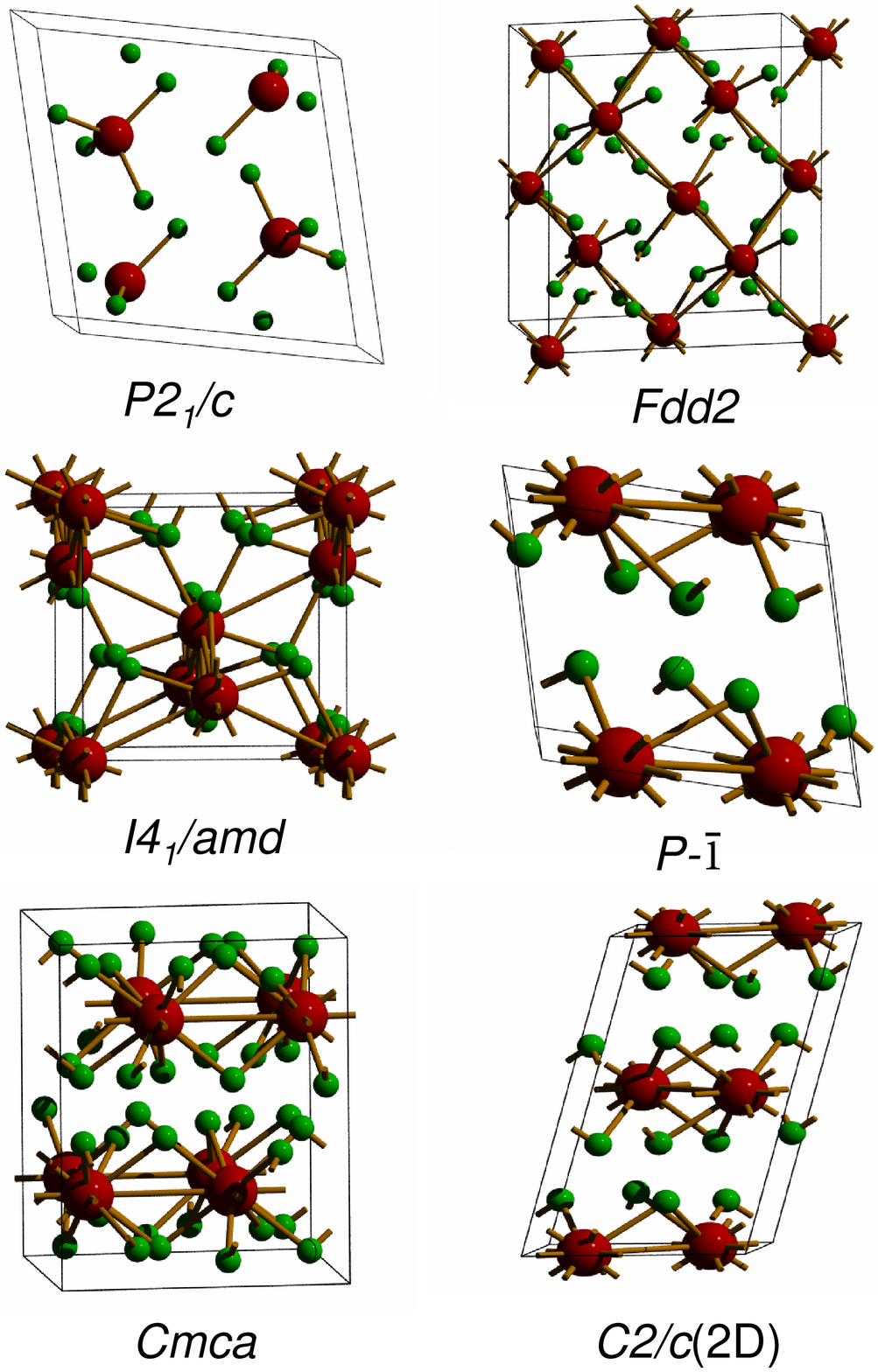}
\caption{(color online) The energetically most favorable structures
computed for SiH$_{4}$ polymorphs at various pressures. }
\end{figure}

Near 50 GPa, there are three other competitive, low-enthalpy
structures with the $C2/c$, $I4_1/amd$, and $I4_1/a$ symmetry. Our
$C2/c$ structure was based on the UI$_4$ arrangement, which as
exhibits a layered structure. However, the $C2/c$ structure
predicted by Pickard and Needs \cite{pick} forms three-dimensional
networks at high pressures. For clarification, we name their
structure as $C2/c$(3D) and our layered structure as $C2/c$(2D).
There are subtle differences in band structures between the $C2/c$
predicted by Yao {\it et al.} \cite{yao} and our $C2/c$(2D). Our
$C2/c$(2D) structure is composed of six-fold coordinated SiH$_6$
octahedra inside a layer, but Si-H bonding between different layers
is absent. It is still dynamically stable up to 250 GPa. However,
the stability of the $C2/c$ phase of Yao {\it et al.} \cite{yao} is
only stable up to 150 GPa. On compression, the six-fold coordinated
SiH$_6$ octahedra are transformed into eight-fold coordinated
SiH$_8$ dodecahedra with a S$_2$H$_2$ bridge-like bonding
arrangement within the layer. Above 110 GPa, the $C2/c$(2D)
structure becomes metallic. The $I4_{1}/amd$ SiH$_{4}$ phase is
composed of eight-fold coordinated SiH$_8$ dodecahedra and every Si
atom shares two H atoms with other Si atoms, as in the $I4_{1}/a$
structure. The $I4_{1}/amd$-type SiH$_4$ is a semimetal between 40
and 70 GPa.

\begin{figure}[tbp]
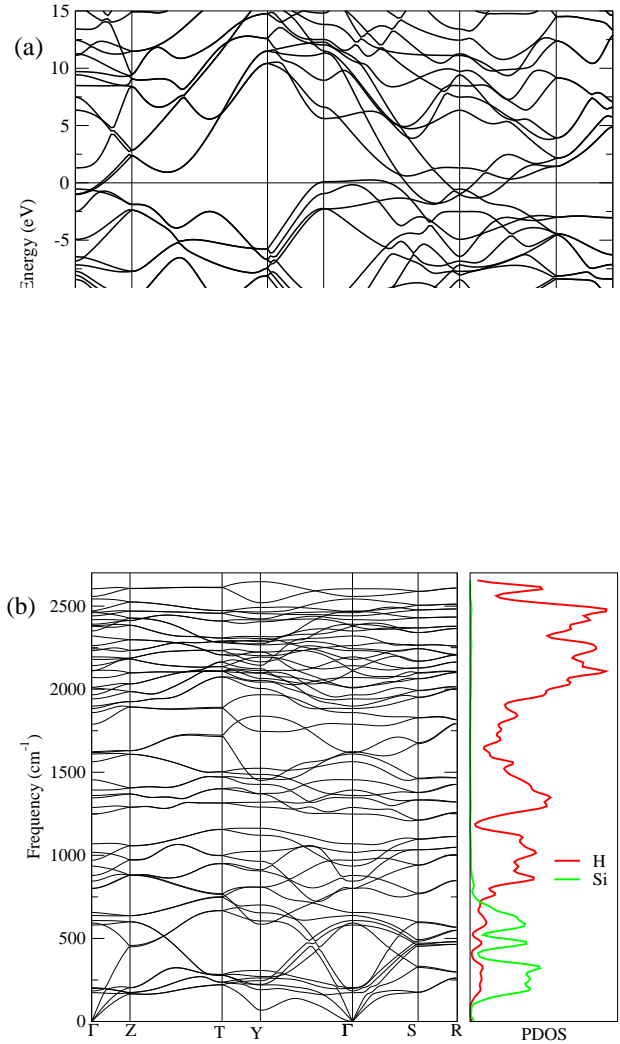

\begin{center}
\includegraphics[width=8cm]{figure3a.eps}
\end{center}

\vspace{2mm}
\begin{center}
\includegraphics[width=8cm]{figure3b.eps}
\end{center}
\caption{(color online) (a) Electronic band structure for the $Cmca$
SiH$_{4}$ phase at 250 GPa. The horizontal line shows the location
of the Fermi level. (b) The phonon dispersion and phonon density of
states (PDOS) projected on Si and H atoms for $Cmca$ SiH$_{4}$ at
250 GPa. }
\end{figure}

The $I4_1/a$ structure was found to have the lowest enthalpy over a
wide pressure range (60 to 220 GPa), consistent with previous
calculations \cite{pick}. In an early powder x-ray diffraction study
\cite{sear}, $I4_{1}/a$ was considered as one of most plausible
structures for the low-temperature phase II. Raman measurements
\cite{chen} indicate that this structure may not exist for SiH$_{4}$
under high pressure and at room temperature. SiH$_{4}$ in the
$I4_1/a$ structure is believed to be stable only at low temperature.
Between 220-270 GPa, a metallic $Cmca$ structure has the lowest
enthalpy. Upon further compression, the $C2/c$(3D) phase possesses
the lowest enthalpy between 270-300 GPa, which confirms the previous
calculations \cite{pick}. The $Cmca$ phase is also a layered
structure that consists of eight-fold coordinated SiH$_{8}$
dodecahedra. In each layer, the Si-H bonding arrangement is the same
as that in the $I4_{1}/a$ structure. Thus, the $Cmca$ phase can be
viewed as the two-dimensional analogue of the $I4_{1}/a$ phase.

In the 60 to 100 GPa range, we found a metallic triclinic structure
with $P\overline{1}$ space group with eight-fold SiH$_{8}$
coordination. It has almost the same enthalpy as the $Cmca$
structure between 60 and 100 GPa. Both of their enthalpies are
within 0.25 eV of the insulating $I4_{1}/a$ structure. As pressure
is increased, the $P\overline{1}$ structure transforms gradually
into the $Cmca$ structure. It is instructive to note that the
$P\overline{1}$ and $Cmca$ phases are metallic over the pressure
regime between 60 GPa and 270 GPa, which is in a good agreement with
recent experiments \cite{chen}. Although the insulating $I4_1/a$
phase has the lowest enthalpy between 60 and 220 GPa, both
$P\overline{1}$ and $Cmca$ are good candidates for metallic phases
in this regime. We thus obtain six energetically favorable
structures for SiH$_{4}$ at high pressures. The atomic arrangements
for each of these structures is shown in Fig. 2. The
$P\overline{1}$, $Cmca$, and $C2/c$(2D) phases have layered
structures and are metallic.

\begin{figure}[tbp]
\includegraphics[width=8cm]{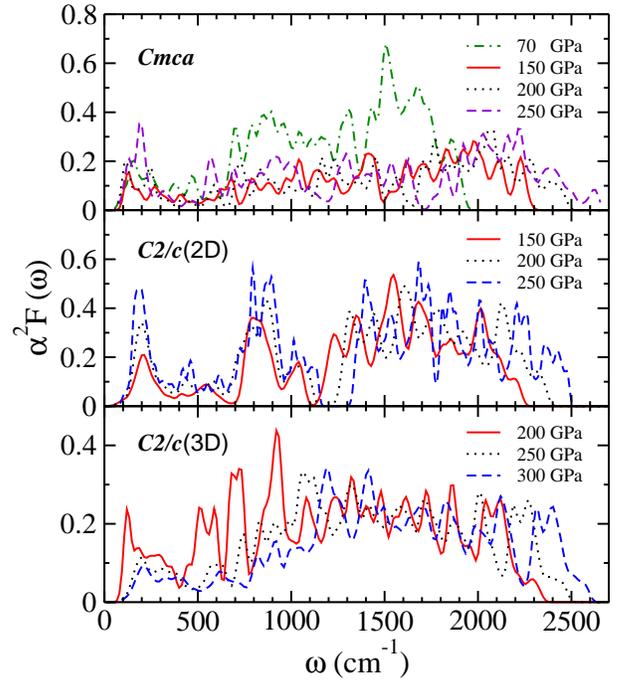}
\caption{(color online) Electron-phonon spectral function
$\alpha^{2}F(\omega)$ vs frequency $\omega$ of metallic SiH$_{4}$
with the $Cmca$, $C2/c$(2D) and $C2/c$(3D) structures at various
pressures. }
\end{figure}

Figure 3(a) shows the  calculated band structure for $Cmca$
SiH$_{4}$ at 250 GPa. It can be seen that the $Cmca$ structure is
metallic. The valence bands cross the Fermi level $E_{F}$ along the
Y$\Gamma$ direction, while the conduction bands cross $E_{F}$ near
the $\Gamma$ point. Upon compression, the conduction band crossing
oss $E_{F}$ at the $\Gamma$ point shifts lower in energy, while the
valence band across $E_{F}$ along the Y$\Gamma$ direction only moves
up slightly in energy. The net effect of the pressure-induced band
shifts is to increase the volume of the Fermi surface and the phase
space for the electron-phonon interaction.

The structural stability of each SiH$_{4}$ phase has been examined
through lattice dynamics calculations. The typical results of the
phonon dispersion and projected phonon density of states for the
$Cmca$ SiH$_{4}$ at 250 GPa are displayed in Fig. 3(b). The $Cmca$
stability is confirmed by the absence of imaginary frequency modes.
There are weak interactions between the Si framework and H atoms
over the whole frequency range. The heavy Si atoms dominate the
low-frequency vibrations, and the light H atoms contribute
significantly to the high-frequency modes. Three separate regions of
bands can be recognized. The modes for the frequencies below 750
cm$^{-1}$ are mainly due to the motions of Si. The bands around 200
cm$^{-1}$ are caused by acoustic phonons. The Si-H-Si bending
vibrations dominate the intermediate-frequency region between 750
and 1200 cm$^{-1}$. At high frequencies above 1200 cm$^{-1}$, the
phonon spectrum belongs to the Si-H bond stretching vibrations.

\begin{figure}[tbp]
\includegraphics[width=\columnwidth]{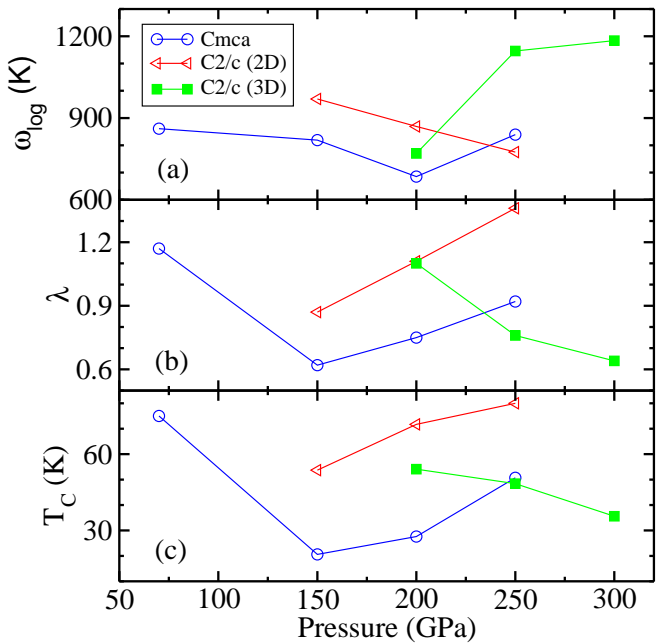}
\caption{(color online)  Calculated (a) logarithmic average phonon
frequency $\omega_{\rm log}$, (b) electron-phonon coupling parameter
$\lambda$, and (c) superconducting transition temperature $T_{c}$ of
SiH$_{4}$ with the $Cmca$, $C2/c$(2D), and $C2/c$(3D) structure as a
function of pressure up to 300 GPa. }
\end{figure}

The electron-phonon spectral function $\alpha^{2}F(\omega)$ is
essential in determining the electron-phonon coupling strength
$\lambda$ and logarithmic average phonon frequency $\omega _{\rm
log}$. We have calculated $\alpha^{2}F(\omega)$ for the metallic
SiH$_{4}$ phases in the pressure range of interest. Figure 4 shows
the results for the $Cmca$, $C2/c$(2D) and $C2/c$(3D) structures
over the pressure range from 70 to 300 GPa. Below 600 cm$^{-1}$ the
major contributions to $\alpha^{2}F(\omega)$ come from the phonon
modes involving Si-Si vibrations, and the remaining part of the
electron-phonon coupling is mainly due to the phonon modes involving
H-H vibrations. The $\alpha^{2}F(\omega)$ on the wide high-energy
side is significantly higher than that on the narrow low-energy
side. Thus the high-energy H-H vibrations dominate the total
$\lambda$ value. Among these metallic structures, the $C2/c$(2D)
phase at 250 GPa has a relatively large $\alpha^{2}F(\omega)$ over
the entire frequency range studied, resulting in a large $\lambda$.

We now examine whether superconductivity in metallic SiH$_{4}$ is
possible, using the $T_{c}$ equation derived by Allen-Dynes
\cite{alle}. In the calculations, we took the Coulomb
pseudopotential $\mu^{*}$ to be 0.1 which was found to reproduce
$T_{c}$ in MgB$_{2}$ \cite{sing}. Figure 5 shows the pressure
dependence of $\omega_{\rm log}$, $\lambda$, and $T_{c}$ for
SiH$_{4}$ with the metallic $Cmca$, $C2/c$(2D) and $C2/c$(3D)
structures. The calculated $\omega_{\rm log}$ decreases [increases]
with pressure for the $C2/c$(2D) [$C2/c$(3D)] structure. However,
$\lambda$ changes with pressure in an opposite sense to $\omega_{\rm
log}$ for both structures. For the $Cmca$ structure, both
$\omega_{\rm log}$ and $\lambda$ do not show a monotonic pressure
dependence. In these metallic phases studied, the variation of
$T_{c}$ with pressure is found to resemble the $\lambda$ behavior.
It is therefore indicated that the pressure effect on $T_{c}$ in
SiH$_{4}$ is primarily controlled by $\lambda$. The $Cmca$ phase has
a $T_{c}$ of 75 K even at 70 GPa. For the $C2/c$(2D) phase, we
calculate a relatively large $T_{c}$ as high as 80 K at 250 GPa. A
decreasing $T_{c}$ from 47.7 K at 200 GPa to 26.1 K at 300 GPa is
obtained for the $C2/c$(3D) phase. The $P\overline{1}$ structure is
also estimated to have a $T_{c}$ of 46.6 K at 70 GPa. The current
results thus suggest new possibilities for exploring high
temperature superconductivity in this hydrogen-rich system.

In summary, we have investigated the structural stability of silane
under pressure up to 300 GPa. The $P2_{1}/c$ phase is confirmed to
be a good candidate for the low-pressure insulating phase. Between
27 and 60 GPa, $Fdd2$ is predicted to be the structure of another
insulating phase. At higher pressures, silane enters the metallic
state having a structure with $P\overline{1}$ symmetry. As pressure
is further increased, the $P\overline{1}$ structure transforms
gradually into the $Cmca$ structure. The three-dimensional $C2/c$
structure is most stable only after 270 GPa. The layered feature of
this material favors metallization and superconductivity. The
relatively high transition temperatures in metallic silane are
mainly attributed to the strong electron-phonon coupling due to the
phonon modes involving H-H vibrations.

We are grateful to J. S. Tse, R. E. Cohen, and S. A. Gramsch  for
discussions and comments. This work was supported by the HKRGC
(402205); the U.S. DOE-BES (DEFG02-02ER34P5), DOE-NNSA
(DEFC03-03NA00144), and NSF (DMR-0205899). X.J.C. wishes to thank
CUHK for kind hospitality during the course of this work. When
revising this manuscript, we were aware of the discovery of
superconductivity in SiH$_{4}$ [M. I. Eremets {\it et al.}, Science
{\bf 319}, 1506 (2008)].

\end{document}